\documentclass[12pt,usenames,dvipsnames]{article}

\usepackage{latexsym}
\usepackage{amssymb,amsfonts,amsmath}
\usepackage{graphicx} 
\usepackage{indentfirst}
\usepackage{bbm}
\usepackage{amssymb}
\usepackage[normalem]{ulem}
\usepackage{verbatim}
\usepackage{amsmath, amsthm, amssymb}
\usepackage{mathrsfs}
\usepackage{hyperref}
\usepackage{amsfonts}
\usepackage{dsfont}
\usepackage{cite}
\usepackage{xcolor}
\usepackage{enumitem}  

\parskip=\medskipamount

\def\a{\alpha}

\def\b{\beta}

\def\g{\gamma}

\def\l{\lambda}

\def\p{\pi}

\def\t{\tau}

\def\z{\zeta}

\def\F{\Phi}

\def\L{\Lambda}

\def\rd{{\rm d}}
\def\ri{{\rm i}}
\def\re{{\rm e}}

\newcommand{\ad}{{\dot{\alpha}}}                           
\newcommand{\bd}{{\dot{\beta}}}                            
\newcommand{\ve}{\varepsilon}                            

\newcommand{\pa}{\partial}                           

\newcommand{\vf}{\varphi}

\newcommand{\be}{\begin{equation}}
\newcommand{\ee}{\end{equation}}
\newcommand{\bea}{\begin{eqnarray}}
\newcommand{\eea}{\end{eqnarray}}
\newcommand{\non}{\nonumber}
\newcommand{\bm}[1]{\mbox{\boldmath$#1$}}

\def\double #1{#1{\hbox{\kern-2pt $#1$}}}

\newif\ifdtup

\newcommand{\bsubeq}{\begin{subequations}}
\newcommand{\esubeq}{\end{subequations}}

\numberwithin{equation}{section}

\newcommand{\sU}{\mathsf{U}}

\newcommand{\mc}{\mathcal}
\newcommand{\mf}{\mathfrak}

\makeatletter
\DeclareFontFamily{OMX}{MnSymbolE}{}
\DeclareSymbolFont{MnLargeSymbols}{OMX}{MnSymbolE}{m}{n}
\SetSymbolFont{MnLargeSymbols}{bold}{OMX}{MnSymbolE}{b}{n}
\DeclareFontShape{OMX}{MnSymbolE}{m}{n}{
    <-6>  MnSymbolE5
   <6-7>  MnSymbolE6
   <7-8>  MnSymbolE7
   <8-9>  MnSymbolE8
   <9-10> MnSymbolE9
  <10-12> MnSymbolE10
  <12->   MnSymbolE12
}{}
\DeclareFontShape{OMX}{MnSymbolE}{b}{n}{
    <-6>  MnSymbolE-Bold5
   <6-7>  MnSymbolE-Bold6
   <7-8>  MnSymbolE-Bold7
   <8-9>  MnSymbolE-Bold8
   <9-10> MnSymbolE-Bold9
  <10-12> MnSymbolE-Bold10
  <12->   MnSymbolE-Bold12
}{}

\let\llangle\@undefined
\let\rrangle\@undefined
\DeclareMathDelimiter{\llangle}{\mathopen}%
                     {MnLargeSymbols}{'164}{MnLargeSymbols}{'164}
\DeclareMathDelimiter{\rrangle}{\mathclose}%
                     {MnLargeSymbols}{'171}{MnLargeSymbols}{'171}
\makeatother

\begin{document}


\begin{titlepage}
\begin{flushright}
December, 2022\\
\end{flushright}
\vspace{5mm}

\begin{center}
{\Large \bf Induced action for superconformal higher-spin multiplets using SCFT techniques}
\end{center}

\begin{center}

{\bf 
Sergei M. Kuzenko, James La Fontaine  and Michael Ponds
}
\vspace{5mm}

\footnotesize{
{\it Department of Physics M013, The University of Western Australia\\
35 Stirling Highway, Crawley W.A. 6009, Australia}}  
\vspace{2mm}
~\\
Email: \texttt{sergei.kuzenko@uwa.edu.au,  michael.ponds@uwa.edu.au}\\
\vspace{2mm}

\end{center}

\begin{abstract}
\baselineskip=14pt
Recently, the interacting ${\cal N}=1$ superconformal higher-spin theory in four dimensions has been proposed within the induced action approach. In this paper we initiate a program of computing perturbative corrections to the corresponding action and explicitly evaluate all quadratic terms. This is achieved by employing standard techniques from superconformal field theory. 
\end{abstract}

\vfill

\vfill
\end{titlepage}

\newpage
\renewcommand{\thefootnote}{\arabic{footnote}}
\setcounter{footnote}{0}

%

\allowdisplaybreaks


\section{Introduction}

Conformal higher-spin (CHS) theory \cite{Tseytlin, Segal} is a rare example of a theory that involves interacting bosonic higher-spin fields, and which also possesses a local Lagrangian formulation.
When defining the corresponding CHS action, there are two routes that one can take: (i) the induced action approach \cite{Tseytlin}; and (ii) the formal operator approach \cite{Segal}.\footnote{For further work related to approach (i) see e.g. \cite{Segal,BJM,BNT,BT,Bonezzi} and for approach (ii) see e.g. \cite{BBJ}. CHS theory has also been recast into the language of BRST/deformation quantisation in \cite{Grigoriev}, see also \cite{GSk}.}
When viewed as an induced action, the crucial ingredient is the coupling $\mc{S}[\varphi,h]$ between complex scalar matter $\varphi$ and  an infinite tower of background CHS fields $h$.
A theory of interacting $\mc{N}=1$ superconformal higher-spin (SCHS) gauge multiplets on a four-dimensional conformally-flat superspace background was recently proposed in \cite{KPR22} as an induced action.
There, the role of $\mc{S}[\varphi,h]$ was played by an action $\mc{S}[\F,H]$ describing chiral matter $\F$ coupled to a tower of SCHS gauge multiplets $H$.

The induced (S)CHS action is formally given by the (logarithmically divergent part of the) 1-loop determinant of the operator characterising the relevant matter coupling. In practice this must be calculated perturbatively in the gauge fields. In the bosonic case, previous approaches to such calculations at the leading-order (i.e. quadratic in $h$) include heat kernel \cite{BJM}, covariant current correlator \cite{BT} and worldline path integral \cite{Bonezzi} techniques. Calculations for the supersymmetric theory have not yet been attempted.

The manifestly (super)conformal formulation of matter couplings provided in \cite{KPR22} allows one to approach the problem from a new angle.
Namely, the $n^{\text{th}}$-order vertices of the (S)CHS action are encoded by the singular contributions to $n$-point correlation functions of primary conserved matter currents. 
The structure of the latter are known to be fixed by the (super)conformal symmetry, allowing one to employ well established results from the (super)conformal field theory ((S)CFT) literature. 
In this letter we exploit this fact to explicitly evaluate all quadratic terms in the induced SCHS action on a Minkowski superspace background, and initiate a program of computing its perturbative corrections.

This paper is organised as follows. In section \ref{secCHS}, as a warm up exercise we describe how the quadratic sector of the induced CHS action may be computed using the set-up of \cite{KPR22} and standard CFT techniques. In section \ref{secSCHS} this method is generalised to a supersymmetric setting, allowing us to compute the quadratic contributions to the induced SCHS action. Some concluding remarks are given in section \ref{secCon}. In appendix \ref{App} we provide an explicit derivation, for the theory under consideration, of the well-known expression for a two-point correlation function of primary tensors fields.

Let us provide some comments on our conventions and notations. Our two-component spinor conventions coincide with those of \cite{BK}. Spinor indices denoted by the same symbol are assumed to be symmetrized over, e.g. 
\begin{align}
U_{\a(m)} V_{\a(n)} = U_{(\a_1 . . .\a_m} V_{\a_{m+1} . . . \a_{m+n})} =\frac{1}{(m+n)!}\big(U_{\a_1 . . .\a_m} V_{\a_{m+1} . . . \a_{m+n}}+\cdots\big)~. \label{convention}
\end{align}
Two distinct spacetime points are differentiated by a subscript, e.g. $x_1^{\a\ad}$ and $x_2^{\a\ad}$. Similarly, in superspace we have $z_1^{A}=(x_1^{\a\ad},\theta_1^{\a},\bar{\theta}^1_{\ad})$ and $z_2^{A}=(x_2^{\a\ad},\theta_2^{\a},\bar{\theta}^2_{\ad})$. In the hope that no confusion arises, we sometimes interchange the positions of the point label and the indices, e.g. $x^1_{\a\ad}$. Similar remarks hold for differential operators at distinct points. 
We often use convention \eqref{convention} to write products of symmetrised objects as, e.g., $(x^1_{\a\ad})^m= x^1_{(\a_1(\ad_1}x^1_{\a_2\ad_2}\cdots x^1_{\a_m)\ad_m)}$. As a rule, when doing so the exponent $m$ will always appear outside a bracket so as not to get misconstrued with a point label.

\section{Quadratic sector of the induced CHS action}\label{secCHS}

The interacting theory of an infinite tower of CHS fields is most easily defined as the action induced by their coupling to a complex scalar field $\varphi$ \cite{Segal, Tseytlin, BJM}. In more detail, one starts with the free matter action 
\begin{align}
\mc{S}_{0}[\vf] = \int \text{d}^4 x \,\bar{\vf} \,\Box \,\vf~,
\end{align}
which is conformally invariant provided $\vf$ is a primary field with conformal weight $1$. This may be coupled to an infinite tower of spin-$s$ conformal fields $h_{\a(s)\ad(s)}$ via a Noether coupling
\begin{align}
\mc{S}[\varphi,h] = \mc{S}_0[\vf]+ \sum_{s=0}^{\infty}\int \text{d}^4 x \, h^{\a(s)\ad(s)}j_{\a(s)\ad(s)}~. \label{CSMatterCoupling}
\end{align}
Both $h_{\a(s)\ad(s)}$ and $j_{\a(s)\ad(s)}$ are real primary tensor fields with conformal weights $2-s$ and $2+s$ respectively, and hence \eqref{CSMatterCoupling} is conformally invariant. 

Currents $j_{\a(s)\ad(s)}$ are bilinear in the matter fields and have the explicit expressions \cite{CDT}
\begin{align}
j_{\a(s)\ad(s)}=\ri^s\sum_{k=0}^{s}(-1)^k\binom{s}{k}^2\pa_{\a\ad}^k\varphi\,\pa_{\a\ad}^{s-k}\bar{\varphi}~.\label{HScur}
\end{align}
They are conserved $\pa^{\b\bd}j_{\b\a(s-1)\bd\ad(s-1)}\approx 0$ on the free equations of motion, $\Box\vf \approx 0$. 
Action \eqref{CSMatterCoupling} is invariant under the gauge transformations 
\begin{align}
\delta h_{\a(s)\ad(s)}&= \pa_{\a\ad}\z_{\a(s-1)\ad(s-1)}+ \mc{O}(h)~, \label{CHSgt}
\end{align}
when accompanied by a suitable linear transformation of the matter field $\vf$. The $\mc{O}(h)$ represents non-Abelian contributions.  In \cite{KPR22}, to formulate a consistent coupling between $\vf$ and the tower of CHS fields, it was useful to introduce a tower of auxiliary gauge fields (which roughly correspond to the trace of the CHS fields used e.g. in \cite{BJM}). These auxiliary fields are not completely necessary in the formulation of $\mc{S}[\vf,h]$ since they may be gauged away, and we have assumed this in the above presentation. 
Gauging away the auxiliary fields alters the gauge transformations of $h_{\a(s)\ad(s)}$ and $\vf$ at order $\mc{O}(h)$, but these changes are not relevant to the analysis of the induced action at quadratic order.

 An effective action $ \Gamma[h]$ may be defined according to
\begin{align}
\re^{\ri \Gamma[h]}=\int\mc{D}\varphi\mc{D}\bar{\varphi}\, \re^{\ri \mc{S}[\varphi,h] }~.\label{EffAct}
\end{align}
As $\mc{S}[\varphi,h]$ is bilinear in $\varphi$, the latter may be integrated out in \eqref{EffAct}, formally giving $\Gamma[h]$. The logarithmically divergent part of $\Gamma[h]$, which we denote schematically as
\begin{align}
\mc{S}_{\text{induced}}[h]:=\Gamma[h]\big|_{\text{UV}}~,
\end{align}
is known as the induced action. It proves to be local, invariant under the gauge transformations \eqref{CHSgt}, and conformally invariant, see e.g. \cite{BJM}. 
Writing $\Gamma[h]=\sum_{n=0}^{\infty}\Gamma^{(n)}[h]$ and expanding both sides of \eqref{EffAct} in a power series in the gauge fields, one finds that
\begin{align}
\Gamma^{(2)}[h]&= \frac{\ri}{2}\sum_{s,s'=0}^{\infty}\int \text{d}^4x_1\int \text{d}^4x_2 \, h^{\a(s)\ad(s)}(x_1)h^{\b(s')\bd(s')}(x_2)\langle j_{\a(s)\ad(s)}(x_1)j_{\b(s')\bd(s')}(x_2)\rangle~.\label{Effquad}
\end{align} 
Here we have made use of the following definition for correlation functions
\begin{align}
\langle \mc{O}_1(x_1)\cdots\mc{O}_n(x_n) \rangle := \frac{1}{Z}\int\mc{D}\varphi\mc{D}\bar{\varphi}\, \re^{\ri \mc{S}_0[\varphi]}\mc{O}_1(x_1)\cdots\mc{O}_n(x_n)~,
\end{align}
with $Z= \int\mc{D}\varphi\mc{D}\bar{\varphi}\, \re^{\ri \mc{S}_0[\varphi]}=\re^{\ri \Gamma^{(0)}}$ and we have also used $\Gamma^{(1)}[h]=0$ which follows from $\langle j_{(s)}\rangle =0$.

Since each current $j_{\a(s)\ad(s)}$ is a primary tensor field with conformal weight $2+s$, only those terms with $s=s'$ contribute to \eqref{Effquad}, and the quadratic part of the induced action reduces to 
\begin{align}
\mc{S}^{(2)}_{\text{induced}}[h] = \frac{\ri}{2}\sum_{s=0}^{\infty}\int \text{d}^4x_1\int \text{d}^4x_2 \, &h^{\a(s)\ad(s)}(x_1)h^{\b(s)\bd(s)}(x_2)\langle j_{\a(s)\ad(s)}(x_1)j_{\b(s)\bd(s)}(x_2)\rangle\big|_{\text{UV}}~. \label{Qind}
\end{align}
The conformal symmetry constrains the above two-point function to take the form \cite{Polyakov}
\begin{align}
\langle j_{\a(s)\ad(s)}(x_1)j_{\b(s)\bd(s)}(x_2)\rangle = c_s\frac{\big(x^{12}_{\a\bd}x^{12}_{\b\ad}\big)^s}{(x_{12}^2)^{2s+2}}~,\qquad x^{12}_{\a\ad}:=x^1_{\a\ad}-x^2_{\a\ad}~, \label{2ptCorrelator}
\end{align}
where $x_{12}^2=x_{12}^ax^{12}_a=-\frac{1}{2}x_{12}^{\a\ad}x^{12}_{\a\ad}$. 

To determine the overall coefficient $c_s$, it is sufficient to compute the totally symmetric part of \eqref{2ptCorrelator}, i.e. $\langle j_{\a(s)\ad(s)}(x_1)j_{\a(s)\ad(s)}(x_2)\rangle$. This can be done by inserting \eqref{HScur}, using Wick's theorem along with the massless two-point correlator (see, e.g., \cite{BogolubovShirkov})
\begin{align}
  \langle \vf(x_1)\bar{\vf}(x_2)\rangle = \frac{1}{4\pi^2\left(x_{12}^2+\ri \ve \right)}\label{2pt}
\end{align}
 and also the identity \eqref{NSida}. After doing this one finds
\begin{align}
c_s = \frac{(-1)^s2^{2s}(2s)!}{16\pi^4}~. \label{NScoeff}
\end{align} 
A simple way to derive \eqref{2pt} is by making use of the proper-time representation for the Feynman propagator
\bea
G_{\rm F} (x_1, x_2) 
=\int \frac{ {\rm d}^4 k }{(2\p)^4}\,
\frac{ {\rm e}^{{\rm i} \,k  \cdot x_{12}}}{k^2 -\ri \ve}
= \int_{0}^{\infty}  \frac{\rd s}{(4 \pi s)^2} \, 
{\rm exp} \Big\{  \frac{\ri \,x_{12}^2 }{4s} - \ve s\Big\} ~, \qquad \ve \to 0^+~.
\eea
For the remainder of the paper we will usually omit the $\ri \ve$ (and have already done so in \eqref{2ptCorrelator}).

We note that, as is well known, the structure of the two-point correlator $\langle j_{s}(x_1)j_{s}(x_2)\rangle$ is fixed by conformal symmetry, a fact which we used in \eqref{2ptCorrelator}. Nevertheless, as an instructive exercise, in appendix \ref{App} we directly compute this correlator and confirm that this result holds true for the free boson. At the same time, this provides a derivation of the coefficients  \eqref{NScoeff} without computing only the totally symmetric part of the correlator.
 
The distribution $(x^2 + \ri 0)^{-\l}$ is complex analytic (with respect to $x^ax_a$) everywhere except for simple poles at $\l=2,3,\dots$, see \cite{GS} for the technical details. 
Introducing the positive infinitesimal regularisation parameter $\tau$, this result may be expressed as follows
\begin{align}
\frac{1}{(x_{12}^2+\ri 0)^{k-\tau}}=\frac{1}{\tau}\frac{-\ri \pi^2}{2^{2(k-2)}(k-1)!(k-2)!}\Box^{k-2}\delta^{4}(x_{12})~+~  
(\text{terms regular in }\tau)~,
\label{Gelfand1}
\end{align}
for $k=2,3,\cdots$.
One may think of $\t$ as the parameter of dimensional regularisation, 
defined by $d= 4-2\t$, which is related to the UV cutoff $\L_{\rm UV}$ as
$1/\t \sim \log \L_{\rm UV}$.

To extract the logarithmically divergent part of \eqref{2ptCorrelator}, we first convert all coordinates in the numerator into total spacetime derivatives and then use \eqref{Gelfand1}. The former step is easily accomplished by employing the two identities (which may be proved by induction on $m$)
\begin{subequations}\label{NSid}
\begin{align}
\frac{(x^{12}_{\a\ad})^m}{(x_{12}^2)^n}&=\left(-\frac{1}{2}\right)^m\frac{(n-m-1)!}{(n-1)!}(\pa_{\a\ad})^m\,\frac{1}{(x_{12}^2)^{n-m}}~,\qquad\qquad\qquad n>m~,\label{NSida}\\
\big[(\pa_{\a\ad})^m,(x^{12}_{\b\bd})^n\big]&=-\sum_{k=1}^{m}2^k\binom{n}{k}\binom{m}{k}k!\big(\ve_{\a\b}\ve_{\ad\bd}\big)^k(\overrightarrow{\pa}_{\a\ad})^{m-k}(x^{12}_{\b\bd})^{n-k}~,\label{NSidb}
\end{align}
\end{subequations}
for positive integers $m$ and $n$ and where $ \partial^{\a\ad} \equiv \partial / \pa x_{\a\ad}^1$. An arrow atop an operator indicates that it acts on everything to its immediate right and continues onwards. The correlator may then be brought into the form
\begin{align}
\langle j_{\a(s)\ad(s)}(x_1)j_{\b(s)\bd(s)}(x_2)\rangle = c_s\sum_{k=0}^{s}\binom{s}{k}^2\frac{k!(k+1)!}{2^{2(s-k)}(2s+1)!}\big(\ve_{\a\b}\ve_{\ad\bd}\big)^k(\pa_{\a\bd}\pa_{\b\ad})^{s-k} \frac{1}{(x_{12}^2)^{k+2}}~.\label{HScor1}
\end{align}

Inserting \eqref{HScor1} into \eqref{Qind}, integrating by parts and using \eqref{Gelfand1} yields
\begin{align}
\mc{S}^{(2)}_{\text{induced}}[h] =\sum_{s=0}^{\infty}\frac{\pi^2c_s}{2^{2s+1}(2s+1)!}\int \text{d}^4x\, h^{\a(s)\ad(s)}\sum_{k=0}^s\binom{s}{k}^2\Box^k\big(\pa_{\a}{}^{\bd}\pa_{\ad}{}^{\b}\big)^{s-k}h_{\a(k)\b(s-k)\ad(k)\bd(s-k)}~.
\end{align}
Finally, we observe that this is equivalent to 
\begin{align}
\mc{S}^{(2)}_{\text{induced}}[h] =\sum_{s=0}^{\infty}\frac{(-1)^s}{32\pi^2(2s+1)}\binom{2s}{s}\int \text{d}^4x\, h^{\a(s)\ad(s)}\mathcal{B}_{\a(s)\ad(s)}(h)~,
\end{align}
where $\mathcal{B}_{\a(s)\ad(s)}(h)$ is the spin-$s$ linearised Bach-tensor (see e.g. \cite{KP19})
\begin{align}
\mathcal{B}_{\a(s)\ad(s)}(h)= \pa_{(\ad_1}{}^{\b_1}\cdots\pa_{\ad_s)}{}^{\b_s}\pa_{(\a_1}{}^{\bd_1}\cdots\pa_{\a_s}{}^{\bd_s}h_{\b_1\dots\b_s)\bd(s)}~.
\end{align}

Therefore, as expected, the quadratic part of the induced action, 
\begin{align}
\mc{S}^{(2)}_{\text{induced}}[h] =\sum_{s=0}^{\infty}\frac{1}{32\pi^2(2s+1)}\binom{2s}{s}\mc{S}_{\text{CHS}}^{(s)}[h]~,
\end{align} 
corresponds to a sum of the well known linearised actions for spin-$s$ conformal gauge fields \cite{FT85}. We note that the action $(-1)^s\mc{S}_{\text{CHS}}^{(s)}[h]$ has the following three equivalent forms
\begin{align}
 \int \text{d}^4x\, h^{\a(s)\ad(s)}\mathcal{B}_{\a(s)\ad(s)}(h)= \int \text{d}^4x\, \mathcal{C}^{\a(2s)}(h)\mathcal{C}_{\a(2s)}(h) = \int \text{d}^4x\, h^{\a(s)\ad(s)}\Box^{s}\Pi_{\perp}^{(s)}h_{\a(s)\ad(s)} ~,
\end{align}
where $\mathcal{C}_{\a(2s)}(h)= \pa_{(\a_1}{}^{\bd_1}\cdots\pa_{\a_s}{}^{\bd_s}h_{\a_{s+1}\dots\a_{2s})\bd(s)}$ is the spin-$s$ linearised Weyl tensor (see e.g. \cite{FL1}) and $\Pi_{\perp}^{(s)}$ is the rank-$s$ transverse projector \cite{BF}.\footnote{The latter has recently been extended to (anti-)de Sitter space in three \cite{AdS3(super)projectors} and four \cite{AdSprojectors} dimensions.}  

\section{Quadratic sector of the induced SCHS action} \label{secSCHS}

In the supersymmetric theory, one considers a coupling between chiral matter and SCHS gauge superfields. The free action for a chiral superfield $\F$  
\begin{align}
\mc{S}_{0}[\F] = \int \text{d}^{4|4} z \,\bar{\F} \, \F~,\qquad \bar{D}_{\ad}\F =0~,
\end{align}
where $\text{d}^{4|4} z = \text{d}^{4} x\text{d}^{2} \theta\text{d}^2\bar{\theta}$, is superconformally invariant provided $\F$ is a primary superfield with conformal weight $1$ and $\sU(1)_R$ charge $-2/3$. Interactions with an infinite tower of background SCHS gauge superfields are introduced via a Noether coupling
\begin{align}
\mc{S}[\F,H] = \mc{S}_0[\F]+\sum_{s=0}^{\infty}\int \text{d}^{4|4} z \, H^{\a(s)\ad(s)}J_{\a(s)\ad(s)} ~. \label{ChiralMatterCoupling}
\end{align} 
Both $H_{\a(s)\ad(s)}$ and $J_{\a(s)\ad(s)}$ are real primary tensor superfields with conformal weights $-s$ and $s+2$ respectively, and hence \eqref{ChiralMatterCoupling} is superconformally invariant. 

Supercurrents $J_{\a(s)\ad(s)}$ have the explicit expressions \cite{KMT}
\begin{align}
J_{\a(s)\ad(s)}=\ri^s\sum_{k=0}^{s}(-1)^k\binom{s}{k}^2\Big(\pa_{\a\ad}^k\Phi\pa_{\a\ad}^{s-k}\bar{\Phi}-\frac{\text{i}}{2}\frac{s-k}{k+1}\pa_{\a\ad}^{k}D_{\a}\Phi\pa_{\a\ad}^{s-k-1}\bar{D}_{\ad}\bar{\Phi}\Big)~,
 \label{supercur}
\end{align}
and are conserved $D^{\b}J_{\b\a(s-1)\ad(s)}\approx \bar{D}^{\bd}J_{\a(s)\ad(s-1)\bd}\approx  0$ on the free equations of motion, $D^2 \F \approx 0 $, where $D^2 = D^{\a}D_{\a}$.  
Action \eqref{ChiralMatterCoupling} is invariant under the gauge transformations 
\begin{align}
\delta H_{\a(s)\ad(s)}&= \bar{D}_{\ad}\L_{\a(s)\ad(s-1)}-D_{\a}\bar{\L}_{\a(s-1)\ad(s)}+ \mc{O}(H)~, \label{SCHSgt}
\end{align}
when accompanied by a suitable linear transformation of the matter field $\F$. The $\mc{O}(H)$ represents non-Abelian contributions.  
Above we have assumed that the auxiliary gauge fields used in \cite{KPR22} to formulate $\mc{S}[\F,H]$ have been gauged away. This alters the gauge transformations of $H_{\a(s)\ad(s)}$ and $\F$ at order $\mc{O}(H)$, which is not relevant to the leading-order analysis in this work. 

 An effective action $ \Gamma[h]$ may be defined following the same procedure as in the previous section, and the quadratic part of the induced action is found to be 
\begin{align}
\mc{S}^{(2)}_{\text{induced}}[H] = \frac{\ri}{2}\sum_{s=0}^{\infty}\int \text{d}^{4|4}z_1\int \text{d}^{4|4}z_2 \, &H^{\a(s)\ad(s)}(z_1)H^{\b(s)\bd(s)}(z_2)\non\\
&\times\langle J_{\a(s)\ad(s)}(z_1)J_{\b(s)\bd(s)}(z_2)\rangle\big|_{\text{UV}}~. \label{SQind}
\end{align}
Superconformal symmetry constrains the two-point function to take the form (see, e.g., \cite{Park1, Osborn1, Park2})
\begin{align}
\langle J_{\a(s)\ad(s)}(z_1)J_{\b(s)\bd(s)}(z_2)\rangle = C_s\frac{\big(x^{\bar{2}1}_{\a\bd}x^{\bar{1}2}_{\b\ad}\big)^s}{~(x_{\bar{1}2}^2x_{\bar{2}1}^2)^{s+1}}~. \label{2ptSCorrelator}
\end{align}
Here we have introduced the two-point building block
\begin{align}
x_{\a\ad}^{\bar{1}2}:= x^{12}_{\a\ad}+2\ri\theta^{12}_{\a}\bar{\theta}^1_{\ad}-2\ri\theta^{2}_{\a}\bar{\theta}^{12}_{\ad}~,\qquad \theta_{\a}^{12}:= \theta_{\a}^1-\theta^2_{\a}~,
\end{align}
which satisfies the identities
\begin{subequations}
\begin{align}
D^1_{\b}x_{\a\ad}^{\bar{1}2} &= 0~,\qquad D^2_{\b}x_{\a\ad}^{\bar{1}2}=-4\ri \ve_{\a\b}\bar{\theta}^{12}_{\ad}~,\\
\bar{D}^2_{\bd}x_{\a\ad}^{\bar{1}2}&=0~,\qquad \bar{D}^1_{\bd}x_{\a\ad}^{\bar{1}2}=\phantom{-}4\ri\ve_{\ad\bd}\theta_{\a}^{12}~.
\end{align}
\end{subequations}
Analogous expressions may be obtained for $x_{\a\ad}^{\bar{2}1}$ by interchanging $1\leftrightarrow 2$. To determine the overall coefficient $C_s$, it is sufficient to compute e.g. the totally symmetric part of \eqref{2ptSCorrelator}, i.e. $\langle J_{\a(s)\ad(s)}(z_1)J_{\a(s)\ad(s)}(z_2)\rangle$. This can be done by inserting \eqref{supercur}, using
\begin{align}
\langle \F(z_1)\bar{\F}(z_2)\rangle = \frac{1}{4\pi^2\left( x_{\bar{2}1}^2+\ri\ve \right)}
\end{align}
 and identity \eqref{Sidc}.\footnote{The general strategy for this is to convert all occurrences of $x_{\a\ad}^{\bar{2}1}$ to $x_{\a\ad}^{\bar{1}2}$. This may be done using the relation $x^{\bar{2}1}_{\a\ad}=-x^{\bar{1}2}_{\a\ad}+4\ri\theta^{12}_{\a}\bar{\theta}^{12}_{\ad}$, from which one can derive the identities $(x^{\bar{2}1}_{\a\ad})^k = (-x^{\bar{1}2}_{\a\ad})^k +  4k\ri \theta^{12}_{\a}\bar{\theta}_{\ad}^{12}(-x^{\bar{1}2}_{\a\ad})^{k-1}$ and $(x_{\bar{2}1}^2)^{-k}=(x_{\bar{1}2}^2)^{-k}-4k\ri x_{\b\bd}^{\bar{1}2}\theta^{\b}_{12}\bar{\theta}^{\bd}_{12}(x_{\bar{1}2}^2)^{-k-1}+4k(k-1)\theta_{12}^2\bar{\theta}_{12}^2(x_{\bar{1}2}^2)^{-k-1} $ for positive integer $k$. Here $\theta_{12}^2:=\theta_{12}^{\a}\theta_{\a}^{12}$.}
After some algebra one finds
\begin{align}
C_s = \frac{2^{2s}(2s+1)!}{16\pi^4 (s+1)}~.
\end{align} 

 To isolate the divergent part of the correlation function \eqref{2ptCorrelator}, we first trade the coordinates in the numerator of \eqref{2ptSCorrelator} for total spinor and vector derivatives. This may be accomplished by use of the identities
\begin{subequations}\label{Sid}
\begin{align}
\frac{1}{(x_{\bar{1}2}^2x_{\bar{2}1}^2)^k}&= \frac{1}{4}D^{\g}_2D^1_{\g}\Big(\frac{\theta_{12}^2}{(x_{\bar{1}2}^2)^{2k}}\Big)+4k(k-1)\frac{\theta_{12}^2\bar{\theta}_{12}^2}{(x_{\bar{1}2}^2)^{2k+1}}~,\label{Sida}\\[5pt]
\big[D^{\g}_2D^1_{\g},\big(x^{\bar{2}1}_{\a\bd}x^{\bar{1}2}_{\b\ad}\big)^s\big]&=4\ri s\Big( \overrightarrow{D}_{\a}^2\bar{\theta}_{\bd}^{12}x_{\b\ad}^{\bar{1}2}- \overrightarrow{D}_{\b}^1\bar{\theta}_{\ad}^{12}x_{\a\bd}^{\bar{2}1}+2\ri s\ve_{\a\b}\ve_{\ad\bd}\bar{\theta}_{12}^2\Big)\big(x^{\bar{2}1}_{\a\bd}x^{\bar{1}2}_{\b\ad}\big)^{s-1}~,\label{Sidb}
\end{align}
in conjunction with the obvious supersymmetric generalisations of \eqref{NSid},
\begin{align}
\frac{(x^{\bar{1}2}_{\a\ad})^m}{(x_{\bar{1}2}^2)^n}&=\left(-\frac{1}{2}\right)^m\frac{(n-m-1)!}{(n-1)!}(\pa_{\a\ad})^m\,\frac{1}{(x_{\bar{1}2}^2)^{n-m}}~,\qquad\qquad\qquad~ n>m~,\label{Sidc}\\
\big[(\pa_{\a\ad})^m,(x^{\bar{1}2}_{\b\bd})^n\big]&=-\sum_{k=1}^{m}2^k\binom{n}{k}\binom{m}{k}k!\big(\ve_{\a\b}\ve_{\ad\bd}\big)^k(\overrightarrow{\pa}_{\a\ad})^{m-k}(x^{\bar{1}2}_{\b\bd})^{n-k}~.\label{Sidd}
\end{align}
\end{subequations}
Eventually, one may bring the correlator \eqref{2ptSCorrelator} into the following form
\begin{align}\label{Scorr2}
\langle J_{\a(s)\ad(s)}(z_1)&J_{\b(s)\bd(s)}(z_2)\rangle = \frac{(-1)^sC_s}{2^{2s-1}(2s+1)!}\sum_{k=0}^{s}2^{2k}\binom{s}{k}^2k!(k+1)!\big(\ve_{\a\b}\ve_{\ad\bd}\big)^k\big(\pa_{\a\bd}\pa_{\b\ad}\big)^{s-k-1}\non\\
&\times\Big\{\frac{1}{8}D^{\g}_2D^1_{\g}\pa_{\a\bd}\pa_{\b\ad}\left(\frac{\theta_{12}^2}{(x_{\bar{1}2}^2)^{k+2}}\right)+(k+2)(s-k)\pa_{\a\bd}\pa_{\b\ad}\left(\frac{\theta_{12}^2\bar{\theta}_{12}^2}{(x_{\bar{1}2}^2)^{k+3}}\right)\non\\[5pt]
\phantom{\times~}-\ri(k&+2)(s-k)D_{\a}^2\pa_{\b\ad}\left(\frac{\bar{\theta}^{12}_{\bd}\theta_{12}^2}{(x_{\bar{1}2}^2)^{k+3}}\right)-\ri(k+2)(s-k)D_{\b}^1\pa_{\a\bd}\left(\frac{\bar{\theta}^{12}_{\ad}\theta_{12}^2}{(x_{\bar{1}2}^2)^{k+3}}\right) \Big\}~.
\end{align}

 Next, to extract the local singular terms, we employ the supersymmetric analogue of \eqref{Gelfand1},
\begin{align}
\frac{\theta_{12}^2}{(x_{\bar{1}2}^2+\ri 0)^{k-\tau}}=\frac{1}{\tau}\frac{-\ri \pi^2}{2^{2(k-2)}(k-1)!(k-2)!}\Box^{k-2}\delta^{6}_+(z_1,z_2)~+~(\text{terms regular in }\tau)~\label{Gelfand2}
\end{align}
for $k=2,3,\cdots$. This may be derived by expanding the Grassmann coordinates and using \eqref{Gelfand1}.
In \eqref{Gelfand2}, $\delta^6_+(z_1,z_2)$ is the chiral superspace delta function, which is related to the full superspace one $\delta^8(z_1-z_2)=\delta^4(x_{12})\theta_{12}^2\bar{\theta}_{12}^2$ via $\delta^6_+(z_1,z_2)=-\frac{1}{4}\bar{D}_1^2\delta^8(z_1-z_2)$ with $\bar{D}_1^2= \bar{D}^1_{\ad}\bar{D}_1^{\ad}$. Inserting \eqref{Scorr2} into \eqref{SQind} and using \eqref{Gelfand2}, the quadratic sector of the induced action is
\begin{align}
&\mc{S}^{(2)}_{\text{induced}}[H] = \sum_{s=0}^{\infty}\frac{(-1)^sC_s\pi^2}{2^{2s+5}(2s+1)!}\int \text{d}^{4|4}z\, H^{\a(s)\ad(s)}\sum_{k=0}^{s}\binom{s}{k}^2\Box^k\big(\pa_{\a}{}^{\bd}\pa_{\ad}{}^{\b}\big)^{s-k-1}\non\\
&\times\Big\{D^{\g}\bar{D}^2D_{\g}\pa_{\a}{}^{\bd}\pa_{\ad}{}^{\b}H_{\a(k)\b(s-k)\ad(k)\bd(s-k)}+8\frac{s-k}{k+1}\Box\pa_{\a}{}^{\bd}\pa_{\ad}{}^{\b}H_{\a(k)\b(s-k)\ad(k)\bd(s-k)}\non\\
&-4\ri\frac{s-k}{k+1}D^{\b}\bar{D}_{\ad}\Box\pa_{\a}{}^{\bd}H_{\a(k)\b(s-k)\ad(k)\bd(s-k)}-4\ri\frac{s-k}{k+1}\bar{D}^{\bd}D_{\a}\Box\pa_{\ad}{}^{\b}H_{\a(k)\b(s-k)\ad(k)\bd(s-k)}\Big\}~.
\end{align}
Finally, we observe that this is equivalent to 
\begin{align}
\mc{S}^{(2)}_{\text{induced}}[H] = \sum_{s=0}^{\infty}\frac{(-1)^{s+1}}{2^{8}\pi^2(s+1)}\binom{2s+2}{s+1}\int \text{d}^{4|4}z\, H^{\a(s)\ad(s)}\mf{B}_{\a(s)\ad(s)}(H)~,
\end{align}
where $\mf{B}_{\a(s)\ad(s)}(H)$ is the superspin-$(s+\frac12)$ linearised super-Bach-tensor \cite{KPR20}
\begin{align}
\mf{B}_{\a(s)\ad(s)}(H)= -\frac{1}{4}\pa_{(\ad_1}{}^{\b_1}\cdots\pa_{\ad_s)}{}^{\b_s}D^{\g}\bar{D}^2D_{(\g}\pa_{\a_1}{}^{\bd_1}\cdots\pa_{\a_s}{}^{\bd_s}H_{\b_1\dots\b_s)\bd(s)}~.
\end{align}

Therefore, as expected, the quadratic part of the induced action, 
\begin{align}
\mc{S}^{(2)}_{\text{induced}}[H] =\sum_{s=0}^{\infty}\frac{1}{2^{8}\pi^2(s+1)}\binom{2s+2}{s+1}\mc{S}_{\text{SCHS}}^{(s)}[H]~,
\end{align} 
corresponds to a sum of the linearised actions for SCHS gauge superfields derived in \cite{KMT}.
The action $(-1)^{s+1}\mc{S}_{\text{SCHS}}^{(s)}[H]$ has the following three equivalent forms
\begin{align}
\int \text{d}^{4|4}z\, H^{\a(s)\ad(s)}\mf{B}_{\a(s)\ad(s)}(H)&=- \int \text{d}^4x\text{d}^2\theta \, \mf{W}^{\a(2s+1)}(H)\mf{W}_{\a(2s+1)}(H) \non\\
&= 2\int \text{d}^{4|4}z\, H^{\a(s)\ad(s)}\Box^{s+1}\bm{\Pi}_{\perp}^{(s)}H_{\a(s)\ad(s)} 
\end{align}
where $\mf{W}_{\a(2s+1)}(H)= -\frac{1}{4}\bar{D}^2\pa_{(\a_1}{}^{\bd_1}\cdots\pa_{\a_s}{}^{\bd_s}D_{\a_{s+1}}H_{\a_{s+2}\dots\a_{2s+1})\bd(s)}$ is the superspin-$(s+\frac 12)$ linearised chiral super-Weyl-tensor and $\bm{\Pi}_{\perp}^{(s)}$ is the  rank-$s$ transverse-linear and transverse-anti-linear $\mc{N}=1$ superprojector \cite{SG}.\footnote{The latter has recently been extended to anti-de Sitter superspace in three \cite{AdS3(super)projectors} and four \cite{AdSuperprojectors} dimensions.} 

\section{Concluding remarks} \label{secCon}

In this letter we have calculated the leading-order contribution to the non-linear SCHS action $\mc{S}_{\text{SCHS}}[H]$ proposed recently in \cite{KPR22}. 
As expected, this contribution corresponds to a sum of the linearised actions for SCHS gauge multiplets derived in \cite{KMT}.
Action $\mc{S}_{\text{SCHS}}[H]$ is realised as the one induced by the coupling $\mc{S}[\F,H]$ of chiral matter to SCHS superfields.  
This coupling is manifestly superconformal, a fact which was heavily exploited in this work to deduce the structure of the correlation functions which encode the induced action. 
 Consequently, this note also provides an alternate method for perturbatively calculating the CHS action, which complements the existing ones \cite{Segal, BJM, BNT, BT, Bonezzi}.

An interesting open problem is to compute the higher-order contributions to the CHS and SCHS actions.
So far, the only progress made in this direction took place within the formal operator setting, where Segal \cite{Segal} derived a closed form expression for the cubic sector of the CHS action.\footnote{Some lower-spin cubic and quartic vertices were also computed in \cite{BNT} in the transverse-traceless gauge. }
  However, the formula provided involves many integration parameters and is quite cumbersome as a result.  
It is possible that a more compact and transparent expression may be obtained by employing the method introduced in this paper. 
In this context, the cubic sector is completely characterised by three-point functions of primary conserved currents of the form $\langle j_{s_1}(x_1)j_{s_2}(x_2)j_{s_3}(x_3) \rangle $ for arbitrary integers $s_i\geq 0$. The general form of correlation functions of this type has been conjectured in \cite{Stanev, Zhiboedov}. In addition to the results of \cite{Stanev, Zhiboedov},  the techniques developed in e.g. \cite{BS4} and references therein would be crucial in such an analysis. In the supersymmetric case, some correlators of this type have been studied in e.g. \cite{BHTM}, though their general structure for arbitrary superspin has not yet been determined.

 Instead of employing known results (where available) from the (S)CFT literature, one could attempt to evaluate the relevant three-point correlators by brute force using diagrammatic/combinatoric techniques, as is done in appendix \ref{App} for the two-point functions. Alternatively, the problem may be approached from an AdS/CFT angle like the non-supersymmetric three-dimensional analysis of \cite{DS13} (so far, a superspace extension of their result has not been worked out). 
 Once the correlators have been computed, the next step is to find a suitable extension of the procedure used in this work to extract the logarithmically divergent part.  These are issues that we plan to address in future work. 


\noindent
{\bf Acknowledgements:}\\
The authors are grateful to Arkady Tseytlin for useful discussions and comments on the manuscript. 
The work of SK and MP is supported in part by the Australian 
Research Council, project No. DP200101944.
MP is grateful to the organisers of the APCTP Workshop ``Higher Spin Gravity and its Applications" (Pohang, South Korea), where part of this work was completed, for the fantastic scientific atmosphere and generous support.

\appendix

\section{Explicit computation of $\langle j_{\a(s)\ad(s)}(x_1)j_{\b(s)\bd(s)}(x_2)\rangle$} \label{App}

In this appendix we explicitly compute the correlator $\langle j_{s}(x_1)j_{s}(x_2)\rangle$ and show that
\begin{align}
\langle j_{\a(s)\ad(s)}(x_1)j_{\b(s)\bd(s)}(x_2)\rangle = \frac{(-1)^s2^{2s}(2s)!}{16\pi^4}\frac{\big(x^{12}_{\a\bd}x^{12}_{\b\ad}\big)^s}{(x_{12}^2)^{2s+2}}~. \label{App1}
\end{align}
To begin, we insert expressions \eqref{HScur} and use Wick's theorem and \eqref{2pt} to deduce
\begin{align}
\langle j_{\a(s)\ad(s)}(x_1)j_{\b(s)\bd(s)}(x_2)\rangle = \frac{1}{16\pi^4}\sum_{k,l=0}^{s}(-1)^{k+l}\binom{s}{k}^2\binom{s}{l}^2\pa_{\a\ad}^k\pa_{\b\bd}^{s-k}\frac{1}{x_{12}^2}\pa_{\a\ad}^{s-l}\pa_{\b\bd}^{l}\frac{1}{x_{12}^2}~. \label{App2}
\end{align}
Next, we convert all $\pa_{\a\ad}$ to $x^{12}_{\a\ad}$ by using \eqref{NSida}, applying the following variant of \eqref{NSidb}
\begin{align}
\big[(\overrightarrow{\pa}_{\a\ad})^m,(x^{12}_{\b\bd})^n\big]=\sum_{k=1}^{m}(-1)^k2^k\binom{n}{k}\binom{m}{k}k!\big(\ve_{\a\b}\ve_{\ad\bd}\big)^k(x^{12}_{\b\bd})^{n-k}(\pa_{\a\ad})^{m-k}~,
\end{align}
and using \eqref{NSida} again. After this one may bring the right-hand-side of \eqref{App2} into the form
\begin{align}
&~~~~~~~~~~~~~~ \frac{2^{2s}}{16\pi^4}\frac{1}{(x_{12}^2)^{2s+2}}\sum_{k,l=0}^{s}(-1)^{k+l}\binom{s}{k}^2\binom{s}{l}^2\Big\{(s+k-l)!(s+l-k)!(x^{12}_{\a\ad}x^{12}_{\b\bd})^s \non\\
&+2\sum_{p=1}^k\binom{s-l}{p}\binom{k}{p}(s+l-k)!(s+k-p-l)!p!(x^{12}_{\a\ad}x^{12}_{\b\bd})^{s-p}(\ve_{\a\b}\ve_{\ad\bd}\, x_{12}^2)^p\\
&+\sum_{p=1}^k\sum_{q=1}^{s-k}\binom{s-l}{p}\binom{l}{q}\binom{s-k}{q}(s+l-k-q)!(s+k-p-l)!p!q!(x^{12}_{\a\ad}x^{12}_{\b\bd})^{s-p-q}(\ve_{\a\b}\ve_{\ad\bd}\, x_{12}^2)^{p+q} \Big\} \non
\end{align}

The first line of this expression may be simplified by use of the identities
\begin{subequations}
\begin{align}
&\sum_{l=0}^s(-1)^l\binom{s}{l}^2(s+k-l)!(s+l-k)! =(-1)^{s+k}s!s!~,\\
&\sum_{k=0}^s\binom{s}{k}^2=\binom{2s}{s} 
\end{align}
whilst the second line may be simplified using
\begin{align}
&\sum_{l=0}^{s}(-1)^l\binom{s}{l}^2\binom{s-l}{p}(s+l-k)!(s+k-l-p)!=(-1)^{s+k}\frac{s!s!}{p!}~,\\
&\sum_{k=p}^{s}\binom{s}{k}^2\binom{k}{p}=\frac{(2s-p)!}{p!}~. 
\end{align}
The third line may be made to look like the second line by using the identities
\begin{align}
&\sum_{l=0}^s (-1)^l\binom{s}{l}^2\binom{s-l}{p}\binom{l}{q}(s+k-p-l)!(s+l-q-k)! = (-1)^{s+k}\frac{s!s!}{p!q!}~,\\
&\sum_{k=1}^{s-1}\sum_{p=1}^{k}\sum_{q=1}^{s-k}\mc{S}_{(k,p,q)}=\sum_{p=2}^s\sum_{q=1}^{p-1}\sum_{k=0}^{s-p}\mc{S}_{(k+q,q,p-q)}~,
\end{align}
\end{subequations}
for an arbitrary summand $\mc{S}_{(k,p,q)}$. The result of these manipulations is
\begin{subequations}\label{AppF}
\begin{align}
\langle j_{s}(x_1)j_{s}(x_2)\rangle &= \frac{(-1)^s2^{2s}(2s)!}{16\pi^4(x_{12}^2)^{2s+2}}\Big((x^{12}_{\a\ad}x^{12}_{\b\bd})^s+\sum_{p=1}^{s}f_{(s,p)}(x^{12}_{\a\ad}x^{12}_{\b\bd})^{s-p}(\ve_{\a\b}\ve_{\ad\bd}\, x_{12}^2)^p\Big)~,\\
f_{(s,p)}&=2\binom{s}{p}^2\binom{2s}{p}^{-1}+\binom{2s}{s}^{-1}\sum_{k=0}^{s-p}\sum_{q=1}^{p-1}\binom{s}{k+q}^2\binom{s-k-q}{p-q}\binom{k+q}{q}~.
\end{align}
\end{subequations}
Using, e.g., Mathematica's { \tt FindSequenceFunction} tool, it may be deduced that $f_{(s,p)}=\binom{s}{p}$, and the final result \eqref{App1} follows from \eqref{AppF} after making use of the identity 
\begin{align}
(x^{12}_{\a\bd}x^{12}_{\b\ad})^s=\sum_{p=0}^s\binom{s}{p}(x^{12}_{\a\ad}x^{12}_{\b\bd})^{s-p}(\ve_{\a\b}\ve_{\ad\bd}\, x_{12}^2)^p~.
\end{align}

\begin{footnotesize}

\end{footnotesize}


\begin{thebibliography}{66}

\bibitem{Tseytlin} 
A.~A.~Tseytlin,
``On limits of superstring in AdS(5) x S**5,''
Theor.\ Math.\ Phys.\  {\bf 133}, 1376 (2002)
[Teor.\ Mat.\ Fiz.\  {\bf 133}, 69 (2002)]
\href{https://arxiv.org/abs/hep-th/0201112}{[arXiv:hep-th/0201112]}.

\bibitem{Segal}
A.~Y.~Segal,
``Conformal higher spin theory,''
Nucl. Phys. B \textbf{664}, 59 (2003)
\href{https://arxiv.org/abs/hep-th/0207212}{[arXiv:hep-th/0207212 [hep-th]]}.

\bibitem{BJM}
X.~Bekaert, E.~Joung and J.~Mourad,
``Effective action in a higher-spin background,''
JHEP \textbf{02}, 048 (2011)
\href{https://arxiv.org/abs/1012.2103}{[arXiv:1012.2103 [hep-th]}].

\bibitem{BNT} 
M.~Beccaria, S.~Nakach and A.~A.~Tseytlin,
``On triviality of S-matrix in conformal higher spin theory,''
JHEP {\bf 1609}, 034 (2016)
\href{https://arxiv.org/abs/1607.06379}{[arXiv:1607.06379 [hep-th]]}. 

\bibitem{BT}
M.~Beccaria and A.~A.~Tseytlin,
``On induced action for conformal higher spins in curved background,''
Nucl. Phys. B \textbf{919}, 359-383 (2017)
\href{https://arxiv.org/abs/1702.00222}{[arXiv:1702.00222 [hep-th]]}.

\bibitem{Bonezzi} 
R.~Bonezzi,
``Induced action for conformal higher spins from worldline path integrals,''
Universe {\bf 3}, no. 3, 64 (2017)
\href{https://arxiv.org/abs/1709.00850}{[arXiv:1709.00850 [hep-th]]}.

\bibitem{BBJ}
T.~Basile, X.~Bekaert and E.~Joung,
``Conformal higher-spin gravity: Linearized spectrum = symmetry algebra,''
JHEP \textbf{11}, 167 (2018)
\href{https://arxiv.org/abs/1808.07728}{[arXiv:1808.07728 [hep-th]]}.

\bibitem{Grigoriev}
M.~Grigoriev,
``Off-shell gauge fields from BRST quantization,''
\href{https://arxiv.org/abs/hep-th/0605089}{[arXiv:hep-th/0605089 [hep-th]]}.

\bibitem{GSk}
M.~Grigoriev and E.~D.~Skvortsov,
``Type-B formal higher spin gravity,''
JHEP \textbf{05}, 138 (2018)
\href{https://arxiv.org/abs/1804.03196}{[arXiv:1804.03196 [hep-th]]}.

 \bibitem{KPR22}
S.~M.~Kuzenko, M.~Ponds, E.~S.~N.~Raptakis,
``Conformal interactions between matter and higher-spin (super)fields,''
 Fortschr. Phys. 2200157 (2022) 
 \href{https://arxiv.org/abs/2208.07783}{[arXiv:2208.07783 [hep-th]]}.
 
 \bibitem{BK} 
I.~L. Buchbinder and S.~M. Kuzenko, {\it Ideas and Methods of Supersymmetry and
Supergravity, Or a Walk Through Superspace},
IOP, Bristol, 1995 (Revised Edition 1998). 

\bibitem{CDT} 
N.~S.~Craigie, V.~K.~Dobrev and I.~T.~Todorov,
``Conformally covariant composite operators in quantum chromodynamics,''
  Annals Phys.\  {\bf 159}, 411 (1985).
  
\bibitem{Polyakov}
A.~M.~Polyakov,
``Conformal symmetry of critical fluctuations,''
JETP Lett. \textbf{12}, 381 (1970).   
  
 \bibitem{BogolubovShirkov}
N.~N.~Bogoliubov and D.~V.~Shirkov,
{\it Introduction to the Theory of Quantized Fields}, (First Russian Edition: 1957),
John Wiley \& Sons, New York, 1980.
  
\bibitem{GS}
I.~M.~Gel'fand and G.~E.~Shilov,
\textit{Generalized Functions, Volume 1: Properties
and Operations}, (First Russian Edition: 1958),
Academic Press, New York, 1964.
  
\bibitem{KP19}
S.~M.~Kuzenko and M.~Ponds,
``Conformal geometry and (super)conformal higher-spin gauge theories,''
JHEP \textbf{05}, 113 (2019)
\href{https://arxiv.org/abs/1902.08010}{[arXiv:1902.08010 [hep-th]]}.  

\bibitem{FT85} 
E.~S.~Fradkin and A.~A.~Tseytlin,  ``Conformal supergravity,''
Phys.\ Rept.\  {\bf 119}, 233 (1985).

\bibitem{FL1} 
E.~S.~Fradkin and V.~Y.~Linetsky,
``Cubic interaction in conformal theory of integer higher-spin fields in 
four dimensional space-time,''  
Phys.\ Lett.\ B {\bf 231}, 97 (1989).

\bibitem{BF} 
R.~E.~Behrends and C.~Fronsdal,
``Fermi decay of higher spin particles,''
Phys.\ Rev.\  {\bf 106}, no. 2, 345 (1957).

\bibitem{AdS3(super)projectors}
 D.~Hutchings, S.~M.~Kuzenko and M.~Ponds, 
``AdS (super)projectors in three dimensions and partial masslessness,'' 
JHEP \textbf{10}, 090 (2021)
\href{https://arxiv.org/abs/2107.12201}{[arXiv:2107.12201 [hep-th]]}.

 \bibitem{AdSprojectors}
S.~M.~Kuzenko and M.~Ponds,
``Spin projection operators in (A)dS and partial masslessness,''
Phys. Lett. B \textbf{800}, 135128 (2020)
\href{https://arxiv.org/abs/1910.10440}{[arXiv:1910.10440 [hep-th]]}.
  
  \bibitem{KMT}
S.~M.~Kuzenko, R.~Manvelyan and S.~Theisen,
``Off-shell superconformal higher spin multiplets in four dimensions,''
JHEP \textbf{07}, 034 (2017)
\href{https://arxiv.org/abs/1701.00682}{[arXiv:1701.00682 [hep-th]]}.

 \bibitem{Park1}
J.~H.~Park,
``N=1 superconformal symmetry in four-dimensions,''
Int. J. Mod. Phys. A \textbf{13}, 1743 (1998)
\href{https://arxiv.org/abs/hep-th/9703191}{[arXiv:hep-th/9703191 [hep-th]]}.
  
\bibitem{Osborn1}
H.~Osborn,
``N=1 superconformal symmetry in four-dimensional quantum field theory,''
Annals Phys. \textbf{272}, 243-294 (1999)
\href{https://arxiv.org/abs/hep-th/9808041}{[arXiv:hep-th/9808041 [hep-th]]}.

\bibitem{Park2}
J.~H.~Park,
``Superconformal symmetry and correlation functions,''
Nucl. Phys. B \textbf{559}, 455 (1999)
\href{https://arxiv.org/abs/hep-th/9903230}{[arXiv:hep-th/9903230 [hep-th]]}.

\bibitem{KPR20}
S.~M.~Kuzenko, M.~Ponds and E.~S.~N.~Raptakis,
``New locally (super)conformal gauge models in Bach-flat backgrounds,''
JHEP \textbf{08}, 068 (2020)
\href{https://arxiv.org/abs/2005.08657}{[arXiv:2005.08657 [hep-th]]}.


\bibitem{SG} 
W.~Siegel and S.~J.~Gates, Jr.,
``Superprojectors,''
Nucl.\ Phys.\ B {\bf 189}, 295 (1981).

\bibitem{AdSuperprojectors}
E.~I.~Buchbinder, D.~Hutchings, S.~M.~Kuzenko and M.~Ponds, 
``AdS superprojectors,'' JHEP {\bf 2104}, 074 (2021)
\href{https://arxiv.org/abs/2101.05524}{[arXiv:2101.05524 [hep-th]]}.

\bibitem{Stanev}
Y.~S.~Stanev,
``Correlation functions of  conserved currents in four dimensional conformal field theory,''
Nucl. Phys. B \textbf{865}, 200 (2012)
\href{https://arxiv.org/abs/1206.5639}{[arXiv:1206.5639 [hep-th]]}.

\bibitem{Zhiboedov}
A.~Zhiboedov,
``A note on three-point functions of conserved currents,''
\href{https://arxiv.org/abs/1206.6370}{[arXiv:1206.6370 [hep-th]]}.

\bibitem{BS4}
E.~I.~Buchbinder and B.~J.~Stone,
``Three-point functions of conserved currents in 3D CFT: general formalism for arbitrary spins,''
Phys. Rev. D \textbf{107}, no.4, 046007 (2023)
\href{https://arxiv.org/abs/2210.13135}{[arXiv:2210.13135 [hep-th]]}.

\bibitem{BHTM}
E.~I.~Buchbinder, J.~Hutomo and G.~Tartaglino-Mazzucchelli,
``Three-point functions of higher-spin supercurrents in 4D ${ N}=1$ superconformal field theory,''
Fortschr. Phys. 2200133 (2022) 
\href{https://arxiv.org/abs/2208.07057}{[arXiv:2208.07057 [hep-th]]}.

\bibitem{DS13}
V.~E.~Didenko and E.~D.~Skvortsov,
``Exact higher-spin symmetry in CFT: all correlators in unbroken Vasiliev theory,''
JHEP \textbf{04}, 158 (2013) \href{https://arxiv.org/abs/1210.7963}{[arXiv:1210.7963 [hep-th]]}.










\end{thebibliography}
\end{document}